\documentstyle[preprint,aps,eqsecnum,epsf]{revtex}

\def\gtsim{$\raisebox{0.6ex}{$>$}\!\!\!\!\!\raisebox{-0.6ex}{$\sim$}\,\,$}

\def\ltsim{$\raisebox{0.6ex}{$<$}\!\!\!\!\!\raisebox{-0.6ex}{$\sim$}\,\,$}

\begin{document}
\draft
\tightenlines

\title {\null\vspace*{-.0cm}\hfill {\small nucl-th/0110004} \\ \vskip
0.8cm
Dissociation of a Heavy Quarkonium at High Temperatures
}

\author{Cheuk-Yin Wong}

\address{Physics Division, Oak Ridge National Laboratory, Oak Ridge,
TN 37831 USA}

\maketitle

\begin{abstract}
We examine three different ways a heavy quarkonium can dissociate at
high temperatures.  The heavy quarkonium can dissociate spontaneously
when it becomes unbound at a temperature above its dissociation
temperature.  Following the recent work of Digal, Petreczky, and Satz,
we calculate the dissociation temperatures of heavy quarkonia taking
into account the angular momentum selection rules and using a
temperature-dependent potential inferred from lattice gauge
calculations.  We find that the selection rules change the
dissociation temperatures substantially for charmonia but only
slightly for bottomia.  A quarkonium system in thermal equilibrium
with the medium can dissociate by thermalization.  The fraction of
quarkonium lying above the dissociation threshold increases as
temperature increases.  A quarkonium can also dissociate by colliding
with light hadrons.  We evaluate the cross sections for the
dissociation of $J/\psi$ and $\Upsilon$ in collision with $\pi$ as a
function of the temperature of the hadron medium, using the
quark-interchange model of Barnes and Swanson.  We find that as the
temperature increases, the threshold energy decreases and the
dissociation cross section increases.

\end{abstract}

\newpage

\section{Introduction}

The suppression of heavy quarkonium production in a quark-gluon plasma
has been a subject of intense interest since the pioneering work of
Matsui and Satz \cite{Mat86}.  Initial insight into the dissociation
temperatures of heavy quarkonium was further provided by Karsch, Mehr,
and Satz \cite{Kar88}.  Recently, Digal, Petreczky, and Satz
\cite{Dig01a,Dig01b} reported theoretical results on the dissociation
temperatures of heavy quarkonia in hadron and quark-gluon plasma
phases.  These are interesting results as they are related to the use
of the suppression of heavy quarkonia production as a signal for the
quark-gluon plasma \cite{Mat86}.

The basic input of Digal $et~al.$ is the temperature dependence of the
$Q$-$\bar Q$ interaction as inferred from lattice gauge calculations
\cite{Kar00}.  We can understand conceptually such a temperature
dependence of the $Q$-$\bar Q$ potential by placing $Q$ and $\bar Q$
as an external source in a QCD medium.  If the temperature of the
medium is zero, the linear confining potential between $Q$ and $\bar
Q$ is the result of the alignment of the color electric fields at
neighboring sites inside the flux tube, in analogy with the alignment of
spins in a ferromagnet (Fig.\ 1$a$).

\vspace*{2.5cm}
\epsfxsize=300pt
\includegraphics{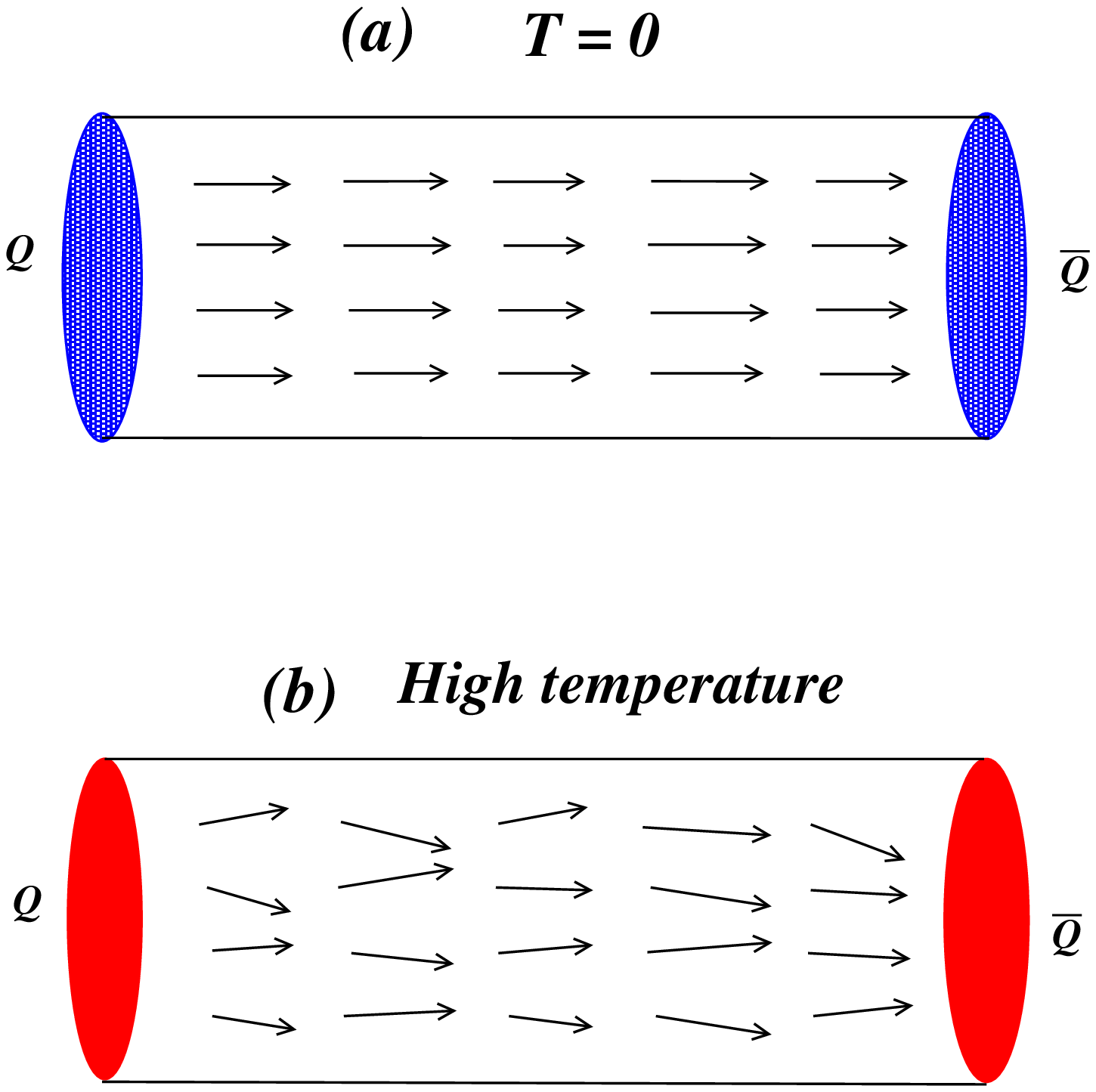}
\vspace*{+4.9cm}\hspace*{1cm}
\begin{minipage}[t]{12cm}
\noindent ({\bf Fig.\ 1$a$}) a schematic configuration of the
color electric field between $Q$ and $\bar Q$ at $T=0$.  ({\bf Fig.\
1$b$}) a possible configuration of the color electric field 
in a hot medium.  Many different configurations of the color electric
fields are possible, with a distribution centered about the mean color
electric field configuration which is aligning along the flux tube.
\end{minipage}
\vskip 4truemm
\noindent 

When the medium is at a high temperature, the gluon and light quark
fields fluctuate and the alignment of the color electric fields due to
the QCD interaction is reduced by the thermal motion for a random
orientation of the color electric fields (Fig.\ 1$b$).  The greater
the temperature, the stronger the tendency for thermal disorientation,
and the weaker the resultant linear interaction between $Q$ and $\bar
Q$.  This tendency continues until deconfinement sets in at the phase
transition temperature $T_c$.

The variation of the $Q$-$\bar Q$ potential gives rise to three
different modes of heavy quarkonium dissociation at high temperatures:
(1) spontaneous dissociation, (2) dissociation by thermal thermalization,
and (3) dissociation by collision with particles in the medium.  We
shall discuss them in turn.

A heavy quarkonium dissociates spontaneously when the binding energy
of the quarkonium relative to a pair of final open charm or open
bottom mesons vanishes.  Utilizing the temperature-dependent potential
for charmonia and bottomia, Digal $et~al.$ found that the dissociation
temperatures for $\psi'$, $\chi_c$, and $\Upsilon'$ in units of $T_c$
are, respectively, $0.1-0.2$, 0.74, and \gtsim 0.83, while the
dissociation temperatures for $J/\psi$, $\Upsilon$, and $\chi_b$ are
above $T_c$ \cite{Dig01a,Dig01b}.

Even at $T=0$ a proper description of the heavy quarkonium state
should be based on a screening potential
\cite{Kar88,Bor89,Din95,Won99}, as the heavy quarkonium becomes a pair
of open charm or open bottom mesons when $r$ becomes very large, due
to the action of dynamical quark pairs.  Previously, a screening
potential was obtained to provide a good description of charmonium
bound states, resonances, and $e^+ e^-$ decay widths at $T=0$
\cite{Won99}.  This screening potential can be used as the starting
point for studying heavy quarkonia at high temperatures.  Using a
potential different from that of Digal $et~al.$ \cite{Dig01a} will
help us find out the sensitivity of the values of dissociation
temperatures on the potentials.  We would also like to include
spin-dependent interactions.  Their inclusion is useful because they
lead to a better description of the energy splittings of the
quarkonium states which affect the dissociation temperatures.
Furthermore, these spin and angular momentum quantum numbers give rise
to selection rules for the final meson states and alter the
dissociation threshold energies and the dissociation temperatures.

The $Q$-$\bar Q$ potential at temperature $T$ refers to the situation
of placing an external quarkonium (at rest) in a medium of temperature
$T$.  The changes of the gluon and light quark fields between the $Q$
and $\bar Q$ due to the temperature of the medium give rise to a
change of the $Q$-$\bar Q$ interaction and the subsequent change of
the binding energy.  The external quarkonium needs not be in thermal
equilibrium with the medium.

The quarkonium placed in a hot medium however will collide with
particles in the medium and become thermalized.  When the quarkonium
reaches thermal equilibrium with the medium, the quarkonium will be in
a mixed state whose occupation probabilities for the different energy
levels will be distributed statistically according to the
Bose-Einstein distribution.  There is a finite probability for this
system to populate states whose energies lie above
their respective dissociation thresholds. This fraction of the
quarkonium system above the thresholds will dissociate into open charm
or open bottom mesons.  Dissociation of this type can be called
dissociation by thermalization.  Because the binding energies of the
quarkonium states decrease as temperature increases, the probability
of dissociation by thermalization also increases as the temperature
increases.

Finally, a heavy quarkonium can also be dissociated by collision with
particles in the medium.  The threshold energies and dissociation
cross sections will be affected by the temperature dependence of the
heavy quarkonium states.  We would like to investigate how these
dissociation cross sections depend on the temperature of the medium,
using the quark-interchange model of Barnes and Swanson \cite{Bar92}
which has been applied successfully to many hadron-hadron
\cite{Bar92,Swa92,Kpi,KN,NN,Bar99} and hadron-(heavy quarkonia)
reactions at $T=0$ \cite{Won00,Won00a,Bar00,Won01}.

In Section II, we describe the Schr\" odinger equation and the
potential used to calculate the energies and the wave functions of
quarkonium states.  We introduce a single-particle
potential with simple spatial and temperature dependence to represent
the results of the lattice gauge calculations.  The single-particle
states are calculated and exhibited in Section III.  We discuss the
selection rules for the spontaneous dissociation of heavy quarkonia in
Section IV. In Section V, the dissociation temperatures are listed and 
compared with previous results of Digal $et~al.$ \cite{Dig01a}.  In
Section VI, we introduce the concept of dissociation by thermalization
and estimate the fraction of the quarkonium system which can
dissociate spontaneously as a function of temperature.  In Section
VII, the cross sections for the dissociation of $\psi$ and $\Upsilon$
in collision with pions are calculated as a function of temperature.
We estimate the survival probability of a heavy quarkonium in
collision with pions in Section VIII. We present our conclusions and
discussions in Section IX.

\section{ Schr\" odinger Equation for Heavy Quarkonium States}

The energy $\epsilon$ of the heavy quarkonium single-particle state
$(Q\bar Q)_{JLS}$ can be obtained by solving the eigenvalue of the
Schr\" odinger equation
\begin{eqnarray}
\label{eq:sch}
\biggl \{ - \nabla \cdot { \hbar^2 \over 2\mu_{12}}\nabla   
+ V_{12}(r,T)
+ \Delta(r,T)
 \biggr \} 
\psi_{JLS} (\bbox{r},T)
 = \epsilon(T) \psi_{JLS} (\bbox{r},T).
\end{eqnarray}
where 
\begin{eqnarray}
\Delta (r,T) = m_1(r,T)+m_2(r,T)
-m_1(\infty,T)-m_2(\infty,T). 
\end{eqnarray}  
We have followed Ref. \cite{Won99} and describe the system as a
two-body system whose properties and masses vary with $r$.  At a
distance $r$\ltsim$R$ where $R \sim 0.8 - 1.0$ fm, the system consists
of a quark and an antiquark, and their masses
$\{m_1(r,T),m_2(r,T)\}=\{m_Q(T), m_{\bar Q}(T)\}$.

The nature of the the two-body system at $r$\gtsim $R$ depends on the
temperature.  If the temperature is below the phase transition
temperature $T_c$ for light quark deconfinement, a heavy quark $Q$ and
antiquark $\bar Q$ separated at large distances will lead to the
production of a light quark pair $\bar q q$ in between and the
subsequent formation of open charm or open bottom mesons $(Q\bar q)$
and $(q\bar Q)$.  At temperatures below $T_c$, the system at
$r$\gtsim$R$ therefore consists of a pair of heavy open flavor mesons,
$(Q\bar q)$ and $(q\bar Q)$, and their masses
$\{m_1(r,T),m_2(r,T)\}=\{M_{Q\bar q}(T),M_{q\bar Q}(T)\}$.  By
including dynamical quarks, the lattice gauge calculations of Karsch
$et~al.$\cite{Kar00} incorporate this change of the configuration at
large distances.  In order to match with the lattice gauge
calculations, the pair of mesons at large distances, $(Q\bar q)$ and
$(q\bar Q)$, should be taken to be the lowest-mass pair, because the
lattice gauge calculation results refer to those of the lowest-energy
states of the system.  The quantity $\mu_{12}$ is the reduced mass.
The interaction $V_{12}$ is the interaction between $Q$ and $\bar Q$
for $r$\ltsim$R$ and between $(Q\bar q)$ and $(q\bar Q)$ for $r$\gtsim
$R$.  As the interaction between mesons has a short range, the
interaction $V_{12}(r,T)$ vanishes as $r$ approaches infinity.

Above $T_c$, light quarks are deconfined and the system is a separated
$Q$ and $\bar Q$.  Thus, $\{m_1(r,T),m_2(r,T)\}=\{m_Q(T),m_{\bar
Q}(T)\}$ and $\Delta(r,T)=0$.  The interaction $V_{12}(r,T)$ also
vanishes as $r$ approaches infinity because the interaction between
$Q$ and $\bar Q$ is screened at large distances in the deconfined
phase.

In Eq.\ (\ref{eq:sch}) the energy $\epsilon(T)$ is measured relative
to the two-body pair at $r\to \infty$.  The quarkonium is bound if
$\epsilon(T)$ is negative.  The quarkonium is unbound and dissociates
spontaneously into two particles if $\epsilon(T)$ is positive, subject
to selection rules which we shall discuss in Section IV.

In the present manuscript, we shall limit our attention to $T \le T_c$
for which the free energy contains contributions only from the
color-singlet component and the extraction of the color-singlet
potential from the lattice gauge calculations is without ambiguity.
For $T$ slightly greater than $T_c$, it is necessary to make
assumptions on the relative fractions of the color-singlet and
color-octet contributions in order to extract the $Q$-$\bar Q$
potential \cite{Dig01b} and the extracted potential there may depend
on the assumed color-singlet and color-octet fractions.

Limiting our attention to $T<T_c$ we note that the mass term $\Delta
(r,T)$ can be approximately represented as
\begin{eqnarray}
\Delta (r,T) \approx [2 m_Q(T)-2M_{Q\bar q}(T)]~\theta (R-r) .
\end{eqnarray}
The meson mass $M_{Q\bar q}(T)$ vary with temperature
\cite{Hay00,Hat01,Bur01}.  Neglecting the kinetic energy of the heavy
quark, the meson mass at temperature $T$ is
\begin{eqnarray} 
M_{Q\bar q}(T)=m_Q(T) + \
\sqrt{ {\bbox{p}}_{\bar q}^2 + [m_{\bar q}(T)]^2}
+ \langle V_{Q\bar q}(T) \rangle
\end{eqnarray}
which leads to 
\begin{eqnarray}
\label{eq:del}
[2m_Q(T)-2M_{Q\bar q}(T)] -
 [2m_Q(T=0)-2M_{Q\bar q}(T=0)]
=-2 \left [ 
\sqrt{ {\bbox{p}}_{\bar q}^2 + [m_{\bar q}(T)]^2}
+ \langle V_{Q\bar q}(T) \rangle \right ]_{T=0}^T
\end{eqnarray}

In the evaluation of the Polyakov loop one obtains the free energy of
the system from the lattice calculations.  This free energy contains
the interaction $V_{12}$ between $Q$ and $\bar Q$ or between $(Q\bar
q)$ and $(q\bar Q)$, depending on the separation $r$.  In addition,
the free energy includes also the term on the right hand side of Eq.\
(\ref{eq:del}) due to the variation of light quark $q$ with
temperature, since the light quark fields are dynamical variables and
they changes the energy of the light quark, and the interaction
between the light quark and the heavy quark $\langle V_{Q\bar
q}(T)\rangle$.  The free energy of the lattice gauge calculations
actually yields an effective interaction $V(r,T)$ which includes
$V_{12}(r,T)$ and the temperature dependence of
$\Delta(r,T)-\Delta(r,0)$:
\begin{eqnarray}
V(r,T)=V_{12}(r,T)+[\Delta(r,T)-\Delta(r,T=0)].
\end{eqnarray}
This effective $Q$-$\bar Q$ interaction $V(r,T)$ vanishes at large
distances.

In terms of this potential $V(r,T)$ extracted from the lattice gauge
calculation, the Schr\" odinger Eq.\ (\ref{eq:sch}) is then
\begin{eqnarray}
\biggl \{ - \nabla \cdot { \hbar^2 \over 2\mu_{12}}\nabla   
+ V(r,T)
+\Delta(r,T=0) \biggr \} 
\psi_{JLS} (\bbox{r},T)
= \epsilon(T) \psi_{JLS} (\bbox{r},T).
\end{eqnarray}

\vspace*{4.1cm}
\epsfxsize=300pt
\includegraphics{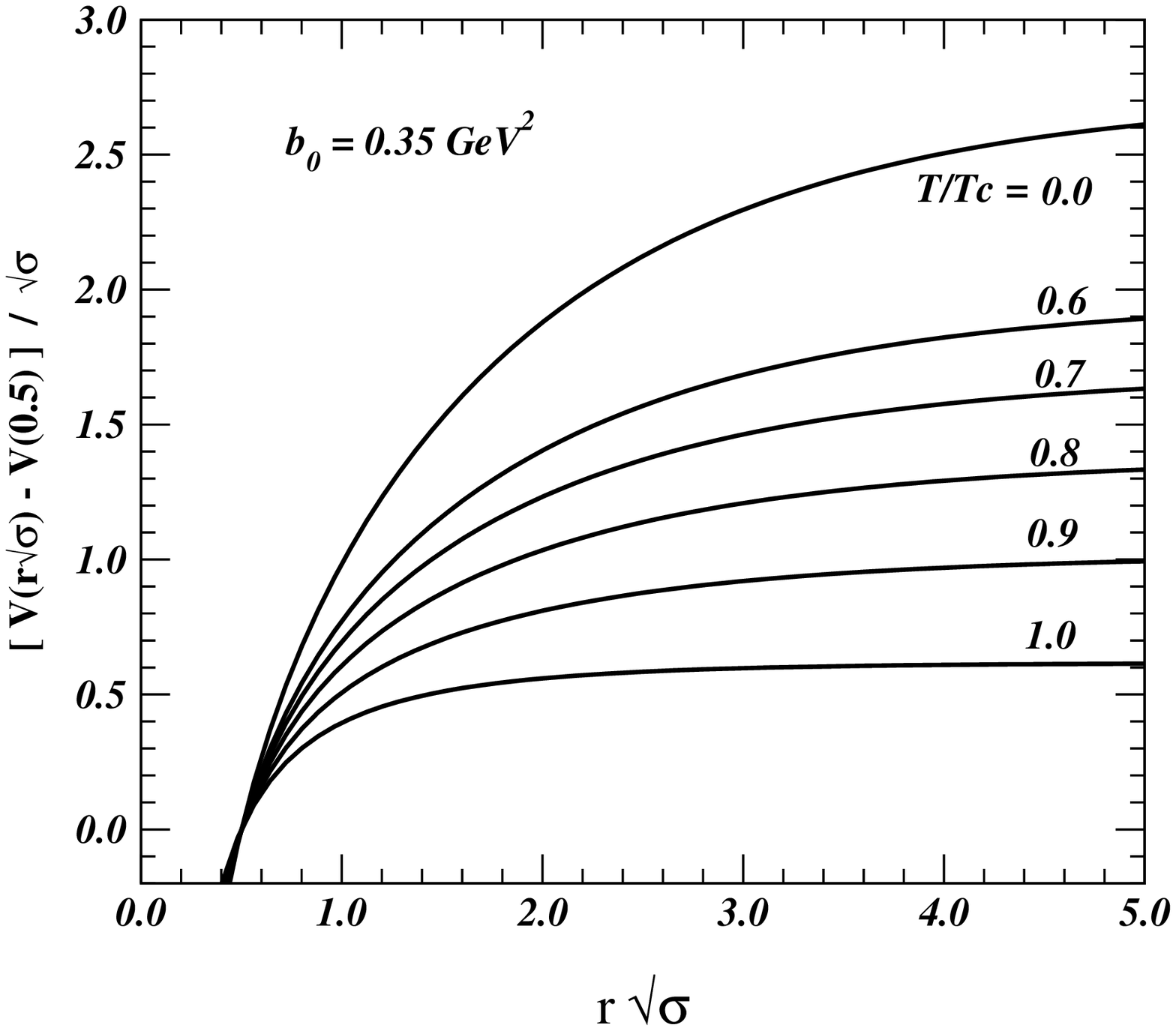}
\vspace*{+6.0cm}\hspace*{1cm}
\begin{minipage}[t]{12cm}
\noindent {\bf Fig.\ 2}.  {Quark-antiquark potential as a function of
$r$ obtained by using Eqs.\ (\ref{eq:pot})-(\ref{eq:pot1}) and
$b_0=0.35$ GeV$^2$. }
\end{minipage}
\vskip 4truemm
\noindent 

Our first task is to obtain a convenient representation of $V(r,T)$
from the results of lattice gauge calculations of Karsch $et~al.$
\cite{Kar00}.  Following Karsch $et~al.$ \cite{Kar88}, we represent
$V(r,T)$ by a Yukawa plus an exponential potential
\cite{Kar88,Bor89,Din95,Won99}
\begin{eqnarray}
\label{eq:pot}
V(r,T)=-{4 \over 3} {\alpha_s e^{-\mu (T) r}\over r}
    -{ b(T) \over \mu(T)} e^{-\mu(T) r}
\end{eqnarray}
where $b(T)$ is the effective string-tension coefficient, $\mu(T)$ is
the effective screening parameter, and the potential is calibrated to
vanish as $r$ approaches infinity.  The results of the lattice
calculations of Ref.\ \cite{Kar00} for $T\le T_c$, which has been
normalized to the Cornell potential \cite{Eic94} at short distances,
can be described by
\begin{eqnarray}
\label{eq:pot1}
b(T)=b_0[1-(T/T_c)^2]\theta(T_c-T),
\end{eqnarray}
and 
\begin{eqnarray}
\label{eq:pot1}
\mu(T)=\mu_0\theta(T_c-T),
\end{eqnarray}
where $b_0=0.35$ GeV$^2$, $\mu_0=0.28$ GeV, and $\theta$ is the step
function.  The value of $\mu_0$ has been fixed to be the same as in
$T=0$, and the value of $b_0$ is close to the value of $b=0.335$
GeV$^2$ obtained earlier at $T=0$ \cite{Won99}.  

\vspace*{4.5cm}
\epsfxsize=300pt
\includegraphics{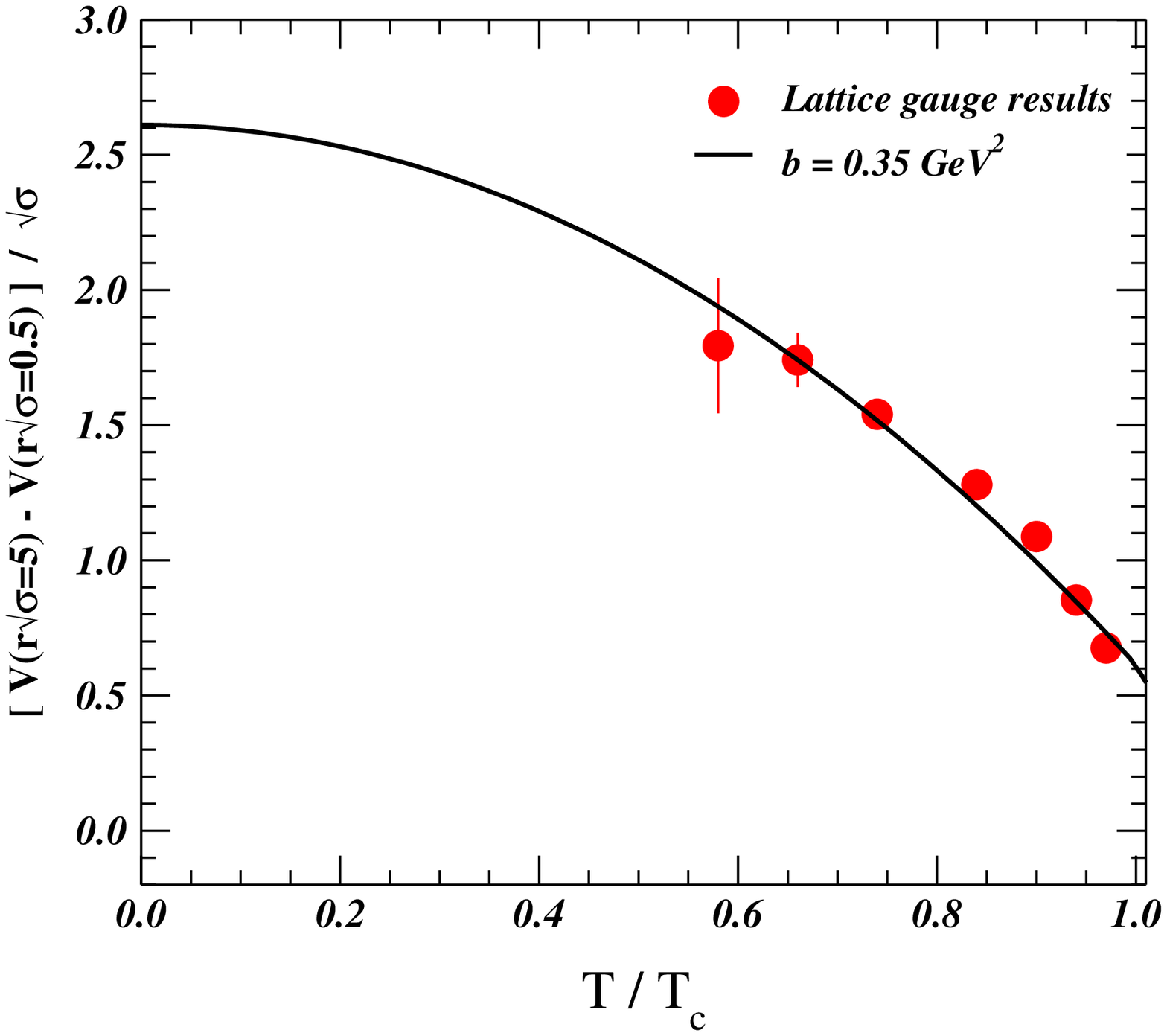}
\vspace*{+5.8cm}\hspace*{1cm}
\begin{minipage}[t]{12cm}
\noindent {\bf Fig.\ 3}.  {The potential difference between
$r\sqrt{\sigma}=5$ and $r\sqrt{\sigma}=0.5$ as a function of the
temperature.  The curve gives the results of Eqs.\ (\ref{eq:pot})
and (\ref{eq:pot1}) with the parameter $b_0=0.35$ GeV$^2$ and the solid
points are lattice gauge results of \cite{Kar00}.}
\end{minipage}
\vskip 4truemm
\noindent 

As a comparison, we plot in Figure 2 the quantity
$[V(r\sqrt{\sigma})-V(0.5)]/ \sqrt{\sigma}$ calculated with the above
potential, Eqs.\ (\ref{eq:pot})-(\ref{eq:pot1}), as a function of
$r\sqrt{\sigma}$ for charmonium at different temperatures. The scale
constant $\sqrt{\sigma}=0.425$ GeV is used to convert distance and
energy to dimensionless numbers as in the lattice gauge calculations
of Ref.\ \cite{Kar00}.  The general features of the spatial and
temperature dependence of the potential in Eqs.\
(\ref{eq:pot})-(\ref{eq:pot1}) agree well with the lattice gauge
results [Fig.\ 6 of \cite{Kar00}].  In particular, there is good
agreement between the variation of the potential difference between
$r\sqrt{\sigma}=5$ and $r\sqrt{\sigma}=0.5$ as a function of the
temperature calculated for charmonium, as shown in Fig.\ 3.

In our calculation, we use a running coupling constant obtained in
another meson spectrum study \cite{Won01},
\begin{eqnarray}
\label{eq:par}
\alpha_s(Q^2)&=&{12 \pi \over (33-2n_f) \ln(A+Q^2/B^2)},
\end{eqnarray}
where $A=10$ and $B=0.31 {\rm ~GeV}$, with $Q$ identified as the mass
of the meson.  The above formula gives $\alpha_s \sim 0.32$ for
charmonium and $\alpha_s \sim 0.24$ for bottomium.  We include, in
addition, the spin-spin, spin-orbit, and tensor interactions,                  
\begin{eqnarray}
\label{eq:pot3}
V_{\rm spin-spin}+V_{\rm spin-orbit}+V_{\rm tensor}=
    {4\over 3}\times{8 \pi \alpha_s \over
3 m_1 m_2 } \bbox{s}_1 \cdot \bbox{s}_2 \left ( {d^3 \over
\pi^{3/2} } \right ) e^{-d^2r^2} 
+ C_{LS} \bbox{L} \cdot \bbox{S} 
+ C_T S_{12},
\end{eqnarray}
where $\bbox{s}_i$ are the spins of the interacting constituents,
$\bbox{L}$ is the orbital angular momentum,
$\bbox{S}=\bbox{s}_1+\bbox{s}_2$, and $S_{12}=\{ 3
(\bbox{\sigma}_1\cdot \hat {\bbox{r}}) (\bbox{\sigma}_2\cdot \hat
{\bbox{r}}) - \bbox{\sigma}_1\cdot\bbox{\sigma}_2 \}$.  For our
calculations, we take $\{C_{LS}, C_T\}= \{34.6, 9.78\}$ MeV for
charmonium, $\{C_{LS}, C_T\}=\{14.25, 2.98\}$ MeV for bottomium.  The
width parameter $d$ is the range of the spin-spin interaction.
Following Godfrey and Isgur \cite{God85}, we use a running width
parameter to include the relativistic effects on the width of the
spin-spin interaction,
\begin{eqnarray}
d^2=\sigma_0^2\left \{{1\over 2} + {1\over 2}\left ( {4 m_1 m_2
\over (m_1+m_2)^2} \right )^4 \right \} 
+\sigma_1^2 \Biggl ( {2 m_1 m_2 \over m_1 +m_2}  \Biggr ) ^2,
\end{eqnarray}
where we obtain the values of $\sigma_0=0.591$ GeV and
$\sigma_1=1.967$ by examining the hyperfine splittings of $\pi$-$\rho$
and $\eta_c$-$J/\psi$.

For $T=0$, $M_{Q \bar q}$ are known from experimental $B$ and $D$
meson masses.  Using the $V(r,T=0)$ extrapolated from lattice gauge
calculations, we vary $m_c$ and $m_b$ and find the zero-temperature
constituent quark masses $m_c = 1.89$ GeV and $m_b=5.22$ GeV which
give a good fit to the heavy quarkonium spectrum (Figs.\ 4 and 5).
These heavy quark constituent mass values are close to those of
$m_c=1.84$ GeV and $m_b=5.18$ GeV in the Cornell potential
\cite{Eic94}, whose short-distance values have been used to normalize
the results of the lattice gauge calculations \cite{Kar00}.

The string tension term in the screening potential Eq.\ (\ref{eq:pot})
needs to have the property that it vanishes at $r\to \infty$, and the
exponential potential has been conveniently chosen to describe such a
property.  It is of interest to compare the exponential potential
$-b_0\exp\{-\mu_0 r\}/\mu_0$ with the standard linear potential
$\sigma r$ at $T=0$.  The effective local string tension, as obtained
by taking $dV/dr$, is $\sigma$ for the linear potential, independent
of $r$.  For the exponential potential, it is $b_0\exp\{-\mu_0 r\}$
which is equal to $b_0$ at $r=0$ and zero at $r \to \infty$.  At the
radius of $J/\psi$, the local string tension is $0.57\, b_0=0.20$
GeV$^2$, which is close to the standard value of $\sigma=0.18$
GeV$^2$, as it should be.  It is therefore reasonable that the
parameter $b_0$ is approximately two times the standard value of
$\sigma$ \cite{Won01}.  The large value of $b_0$ in comparison with
$\sigma$ was noted earlier by Ding $et~al.$ \cite{Din95}.
 
\section{Heavy quarkonium single-particle states}
\label{sec:sps}

The eigenvalues of the Hamiltonian can be obtained by matrix
diagonalization using a set of nonorthogonal Gaussian basis states
with different widths, as described in Ref.\ \cite{Won01}.  For the
construction of the Hamiltonian matrix, the matrix elements of the
Yukawa potential and exponential potential between Gaussian basis
states are
\begin{eqnarray}
\langle i| e^{-\mu r}|j\rangle
= \left ({ 2 \sqrt{ij}\over i+j}\right ) ^ {l+3/2}
{ 2^{l+3/2} (l+1)! \over \sqrt{\pi}} \exp\{ {\mu^2 \over 4(i+j)\beta^2}
 \} U \left (2l+5, {\mu \over \sqrt{(i+j)}\beta} \right ),
\end{eqnarray}
and
\begin{eqnarray}
\langle i| {e^{-\mu r}\over r}|j\rangle
=  \left ({ 2 \sqrt{ij}\over i+j}\right ) ^ {l+3/2}
{ 4 \pi \beta \sqrt{i+j} (2l+1)! \over (2\pi)^{3/2}(2l+1)!!} 
\exp\{ {\mu^2 \over 4(i+j)\beta^2}
 \} U \left (2l+{3\over 2}, {\mu \over \sqrt{(i+j)}\beta} \right ),
\end{eqnarray}
where $U$ is the cylinder parabolic function, $\beta$ is the width
parameter of the wave function basis, and we have used the notation
and normalization of Ref.\ \cite{Won01}.  The matrix elements of the
kinetic energy operator are
\begin{eqnarray}
\langle i |- \nabla \cdot { \hbar^2 \over 2\mu_{12}(\bbox{r})}\nabla
|j \rangle=T_{ij>}+T_{ij<}
\end{eqnarray} 
where
\begin{eqnarray}
T_{ij>}={\hbar^2\over 2 \mu_>} \left ( {2 \sqrt{ij} \over i+j}\right )^{l+3/2}
(i+j) \beta^2 &&\biggl \{  {l\over (2l+1)!!} \left [ -(2l+1)2^l I_l(z)
+2^{l+1} I_{l+1}(z)\right ]
\nonumber\\
&&+{ij \over (i+j)^2} {(2l+3)!! - 2^{l+2} I_{l+2}(z) \over (2l+1)!!}
  \biggr \},
\end{eqnarray}
\begin{eqnarray}
T_{ij<}={\hbar^2\over 2 \mu_<} \left ( {2 \sqrt{ij} \over i+j}\right )^{l+3/2}
(i+j) \beta^2 &&\biggl \{ {l\over (2l+1)!!} \left [ (2l+1)2^l I_l(z)-2
^{l+1} I_{l+1}(z)\right ]
\nonumber\\
&&+{ij \over (i+j)^2} {2^{l+2} I_{l+2}(z) \over (2l+1)!!}
\biggr \},
\end{eqnarray}
\begin{eqnarray}
z=\sqrt{{i+j\over 2}}\beta R,
\end{eqnarray}
\begin{eqnarray}
I_n(z)={2\over \sqrt{\pi}} \int_0^z t^{2n}e^{-t^2} dt=(-1)^n\left \{
\left (\partial \over d\lambda \right )^n [\lambda^{-1/2}
{\rm erf}(\lambda^{1/2} z)] \right \}_{\lambda=1},
\end{eqnarray}
and $\mu_>$ and $\mu_<$ are the reduced masses in the region $r>R$ and
$r<R$ respectively.

After the Hamiltonian matrix is constructed using these matrix
elements, the single-particle state energies and wave functions can be
obtained by matrix diagonalization.

In Figs.\ 4 and 5, we show charmonium and bottomium single-particle
states as a function of temperature. On the left panel of each figure,
the experimental single-particle energies at $T$=0 are also shown for
comparison.  The single-particle energies $\{ \epsilon_i \}$ rise with
increasing temperatures.  At a certain temperature, the energy of a
single-particle state will rise above its threshold for dissociation.
But what is its dissociation threshold?

\section{ Selection rules for spontaneous dissociation of heavy quarkonium}
\label{sec:sps1}

In considering the dissociation below $T_c$, it is necessary to find
the selection rules for the spontaneous dissociation of a heavy
quarkonium state with initial quantum numbers $J$, $L_i$, and $S_i$
into two mesons with a total spin $S$ and a relative orbital angular
momentum $L$.  We shall only discuss the selection rules for
charmonium states, as similar rules apply in the case of bottomium
states.  We limit our attention to final states consisting of a $D$ or
$D^*$ with an antiparticle $\bar D$ or $\bar D^*$.  The parity of a
quarkonium state is $(-1)^{L_i+1}$.  On the other hand, the parity of
a pair of dissociated open charm mesons is $(-1)^L$.  Parity
conservation requires $\Delta L=|L-L_i|=1$ in this dissociation.
Hence, $J/\psi$ and $\psi'$ will dissociate into a pair of open charm
mesons with $L=1$, while $\chi$ states will dissociate into a pair of
open charm mesons with $L=0$ or 2.

\vspace*{3.5cm}
\epsfxsize=300pt
\includegraphics{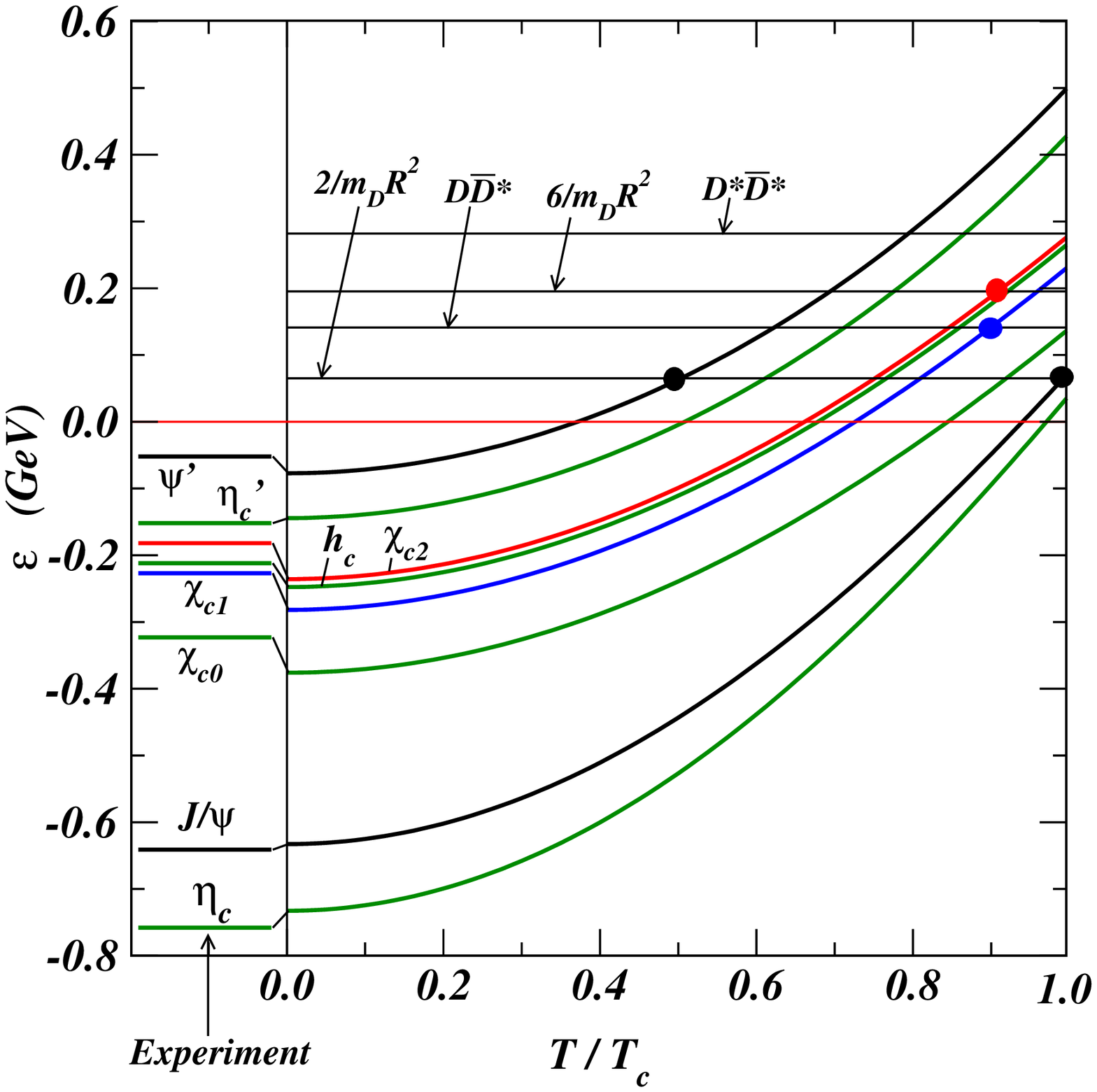}
\vspace*{+8.8cm}\hspace*{1cm}
\begin{minipage}[t]{12cm}
\noindent {\bf Fig.\ 4}.  {Charmonium single-particle states as a
function of temperature.  The threshold energies are indicated as
horizontal lines.  The solid circles indicate the locations of the
dissociation temperatures for states with significant decay branching
fractions into $J/\psi$. }
\end{minipage}
\vskip 4truemm
\noindent 

The threshold for spontaneous dissociation of $J/\psi$ and $\psi'$ is
therefore $M(D)+M(\bar D) + 2/2\mu_{D\bar D} R^2$ where $\mu_{D\bar
D}$ is the reduced mass of the final two-meson system and the term
$2/2\mu_{D\bar D} R^2$ represents the $L=1$ centrifugal barrier height
at the distance $R$ at which $D$ and $\bar D$ become on the
mass-shell.  The reduced mass $\mu_{D \bar D}$ is equal to $m_D/2$.
For numerical purposes, we use $R=0.8$ fm.

Consider now the dissociation of $\chi_{c2}$.  If the final orbital
angular momentum of the two-meson state is $L=0$, then because
$J=L+S=2$, the final two-meson system must have $S=2$ and the two
mesons must each have $S=1$.  The threshold energy for the
dissociation of $\chi_{c2}$ into two mesons in the $L=0$ state is
$M(D^*)+M(\bar D^*)$.  On the other hand, if the final state has
$L=2$, then any spin combination of the final meson is allowed, and
the lowest threshold for the dissociation of $\chi_{c2}$ into two
mesons in the $L=2$ state is $M(D)+M(\bar D)+6/2\mu_{D\bar D} R^2$.

Consider next the dissociation of $\chi_{c1}$.  If the final state has
$L=0$, then because $J=L+S=1$, the final meson states must have
$S=1$. The threshold energy for the dissociation of $\chi_{c1}$ into
two mesons in the $L=0$ state is $M(D)+M(\bar D^*)$. The dissociation
of $\chi_1$ into $D+\bar D$ is not allowed.

\vspace*{3.5cm}
\epsfxsize=300pt
\includegraphics{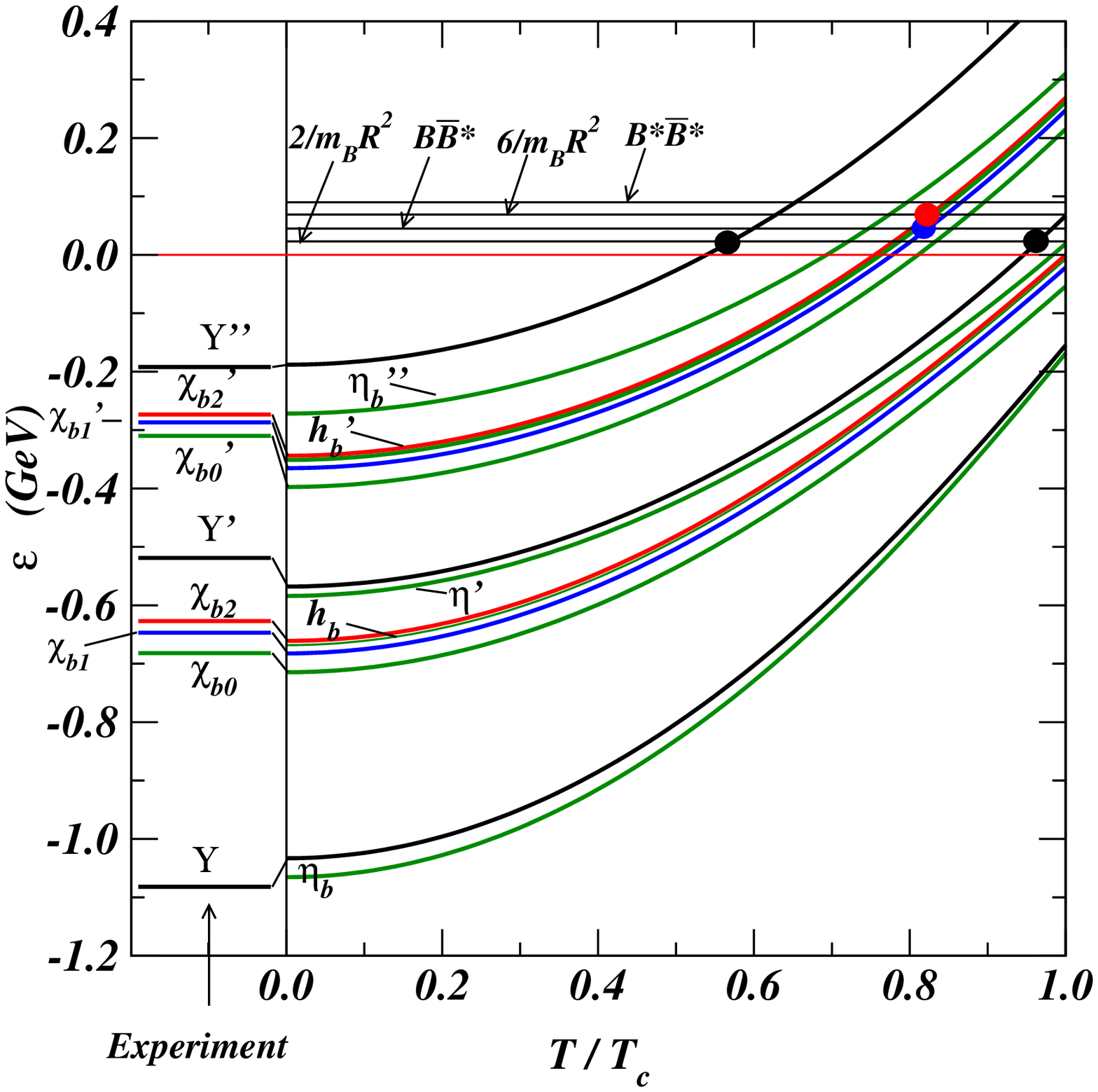}
\vspace*{+8.8cm}\hspace*{1cm}
\begin{minipage}[t]{12cm}
\noindent {\bf Fig.\ 5}.  {Bottomium single-particle energies as a
function of temperature.  The solid circles indicate the locations of
the dissociation temperatures for states with significant decay
branching fractions into $\Upsilon$. }
\end{minipage}
\vskip 4truemm
\noindent 

One can obtain the selection rules for $\eta_c$, $\chi_{c0}$, and
$h_c$ in a similar way.  We list the selection rules for the $S$-wave
charmonium states ($J/\psi$, $\psi'$, and $\eta_c$) in Table I, and
the $P$-wave charmonium states ($\chi_{cJ}$ and $h_c$) in Table II.

\vskip 0.5cm \centerline{Table I.  Selection rules for the
dissociation of $S$-wave charmonia.  } {\vskip 0.0cm\hskip 5cm
\begin{tabular}{|c|c|c|c|} \hline
{\rm Initial Heavy Charmonium} 	& ~{\rm Final~Mesons}  & ~~$L$~~~&
{\rm Threshold}     \\
\hline
$J/\psi,~~\psi'$ &$D  +\bar D$  & 1 & $M(D )+M(\bar D  )+2/2\mu_{D\bar
D} R^2$ \\  
                                                        \cline{2-4}
             &$D  +\bar D^*$ & 1 & $M(D  )+M(\bar D^*)
+2/2\mu_{D \bar D^*}R^2$ \\  
                                                        \cline{2-4}
             &$D^*+\bar D^*$& 1 & $M(D^*)
+M(\bar D^*)+2/2\mu_{D^*\bar D^*}R^2$ \\  
                                                        \hline
$\eta_c$  &$D  +\bar D  $& 1 & {\rm ~not~allowed}                    \\  
                                                        \cline{2-4}
             &$D  +\bar D^*$ & 1 & $M(D  )+M(\bar D^*)
+2/2\mu_{D \bar D^*}R^2$ \\  
                                                        \cline{2-4}
             &$D^*+\bar D^*$& 1 & $M(D^*)
+M(\bar D^*)+2/2\mu_{D^*\bar D^*}R^2$ \\  
                                                        \hline
\end{tabular}
}
\vskip 0.6cm


\vskip 0.0cm \centerline{Table II.  Selection rules for the
dissociation of $P$-wave
charmonia. } {\vskip 0.0cm\hskip 5cm
\begin{tabular}{|c|c|c|c|} \hline
{\rm Initial Heavy Charmonium} 	& ~{\rm Final~Mesons}  & ~~$L$~~~&
{\rm Threshold}     \\
\hline
$\chi_{c2}$  &$D  +\bar D  $& 0 & {\rm ~not~allowed}                    \\  
                                                        \cline{3-4}   
             &              & 2 &$ M(D  )+M(\bar D  )
+6/2\mu_{D\bar D}R^2$ \\  
                                                        \cline{2-4}
             &$D  +\bar D^*$& 0 & {\rm ~not~allowed}                    \\  
                                                        \cline{3-4}   
             &              & 2 &$ M(D  )+M(\bar D^*)
+6/2\mu_{D\bar D^*}R^2$ \\  
                                                        \cline{2-4}
             &$D^*+\bar D^*$& 0 &$ M(D^*)  + M(\bar D^*)$        \\
                                                        \cline{3-4}   
             &              & 2 &$ M(D^*)
+M(\bar D^*)+6/2\mu_{D^*\bar D^*}R^2$ \\  
                                                        \hline

$\chi_{c1}$  &$D  +\bar D  $& 0 & {\rm ~not~allowed}             \\  
                                                        \cline{3-4}   
             &              & 2 & {\rm ~not~allowed}             \\  
                                                        \cline{2-4}   
             &$D  +\bar D^*$& 0 &$  M(D  )+M(\bar D^*)$          \\  
                                                        \cline{3-4}   
             &              & 2 &$ M(D  )+M(\bar D^*)
+6/2\mu_{D\bar D^*}R^2$ \\  
                                                        \cline{2-4}
             &$D^*+\bar D^*$& 0 &$ M(D^*)+M(\bar D^*)$                    \\  
                                                        \cline{3-4}   
             &              & 2 &$ M(D^*)+M(\bar D^*)
+6/2\mu_{D^* \bar D^*}R^2$ \\  
                                                        \hline
$\chi_{c0}$  &$D  +\bar D  $& 0 & $M(D)+M(\bar D)$               \\  
                                                        \cline{3-4}   
             &              & 2 & {\rm ~not~allowed}             \\  
                                                        \cline{2-4}   
             &$D  +\bar D^*$& 0 & {\rm ~not~allowed}             \\  
                                                        \cline{3-4}   
             &              & 2 & {\rm ~not~allowed}             \\  
                                                        \cline{2-4}
             &$D^*+\bar D^*$& 0 &$ M(D^*)+M(\bar D^*)$                    \\  
                                                        \cline{3-4}   
             &              & 2 &$ M(D^*)+M(\bar D^*)
+6/2\mu_{D^* \bar D^*}R^2$ \\  
                                                        \hline

$h_c$        &$D  +\bar D  $& 0 & {\rm ~not~allowed}                    \\  
                                                        \cline{3-4}   
             &              & 2 & {\rm ~not~allowed}             \\  
                                                        \cline{2-4}   
             &$D  +\bar D^*$& 0 &  $ M(D)+M(\bar D^*)$  \\  
                                                        \cline{3-4}   
             &              & 2 &  $ M(D)+M(\bar D^*)
+6/2\mu_{D^* \bar D^*}R^2$ \\  
                                                        \cline{2-4}
             &$D^*+\bar D^*$& 0 &$ M(D^*)+M(\bar D^*)$                    \\  
                                                        \cline{3-4}   
             &              & 2 &$ M(D^*)+M(\bar D^*)
+6/2\mu_{D^* \bar D^*}R^2$ \\  
                                                        \hline
\end{tabular}
}
\vskip 0.6cm

Similar selection rules can be written down for bottomium dissociation
by replacing $J/\psi$ by $\Upsilon$, $\psi'$ by $\Upsilon'$, $\chi_c$
by $\chi_b$, $h_c$ by $h_b$, and $D$ by $B$ in Tables I and II.

\section{Dissociation temperatures for charmonia and bottomia}

Fig.\ 4 shows the charmonium single-particle states as a function of
temperature.  The dissociation temperatures of different quarkonium
states can be determined by plotting the threshold quantities and the
state energies of the quarkonia.  The points of intercept, as
indicated by solid circles in Fig.\ 4, give the positions of the
dissociation temperatures.  Because the observed $J/\psi$ receives
little feeding from the $\chi_{c0}$ and $h_c$ states, and $\eta_c$ is
not observed in dilepton decays, we shall not include the dissociation
temperatures of $\chi_{c0}$, $h_c$, and $\eta_c$ (and similarly
$\chi_{b0}$, $h_b$, and $\eta_b$) in our consideration.  Fig.\ 5 gives
similar temperature dependence of the bottomium single-particle states
as a function of temperature.

\vspace*{0.5cm}
\centerline{Table III.  The dissociation temperatures $T_d$ in
units of $T_c$ for various heavy quarkonia}
{\vskip 0.0cm\hskip 5cm
\begin{tabular}{|c|c|c|c|c||c|c|c|c|c|c|c|} \hline
\vspace*{-0.0cm}
{\rm Heavy           }& $~~\psi'~~$ &$~~\chi_{c2}~~$ 
                      & $~~\chi_{c1}~~$ & $~~J/\psi~~$ 
                      & $~~\Upsilon''~$ & $~~\chi_{b2}'~$ 
                      & $~~\chi_{b1}'~~$
                      & $~~\Upsilon'~~$ & $~~\chi_{b2}~~$ 
                      & $~~\chi_{b1}~~$ & $~~\Upsilon~~$
                        \\
{\rm Quarkonium}     &  &  &  & 
		     &  &  &
                     &  &  &  &  \\
\hline
 $T_d/T_c$             & 0.50 & 0.91 & 0.90&  0.99
                   & 0.57 & 0.82  & 0.82  
                   & 0.96 & $>$ 1.00 & $>$ 1.00  & $>$1.00   \\
\hline  
\vspace*{-0.0cm}
$T_d/T_c$             & 0.37 & 0.66 & 0.72&  0.94
                            & 0.54 & 0.75  & 0.78                    
                   & 0.95 & 1.00 & $>$1.00  & $>$1.00   \\
\vspace*{-0.0cm}
(Selection Rules      &  &  &  & 
		     &  &  &
                     &  &  &  &  \\
not invoked)          &  &  &  & 
		     &  &  &
                     &  &  &  &  \\
\hline
\vspace*{-0.0cm}
 $T_d/T_c$         & 0.1-0.2 & 0.74 & 0.74& 1.10
                   & 0.75 & 0.83  & 0.83  
                   & 1.10 & 1.13 & 1.13  & 2.31   \\
(Digal $et~al.$)     &  &  &  & 
		     &  &  &
                     &  &  &  &  \\
\hline
\end{tabular}
}
\vskip 0.6cm

We list the dissociation temperatures of charmonia and bottomia in
Table III.  If the dissociation threshold is just $M(D)$+$M(\bar D)$,
the dissociation temperatures in units of $T_c$ for $\psi'$,
$\chi_{c2}$, $\chi_{c1}$, and $J/\psi$ in units of $T_c$ would be
0.37, 0.66, 0.72, and 0.94, as listed in the second row of Table III.
This is in rough agreement with the results of Digal $et~al.$ which
give the values of 0.1-0.2, 0.74, and 1.1 for $\psi'$, $\chi_c$, and
$J/\psi$ respectively.  One of the main reasons for the observed
differences is that the locations of the dissociation temperatures
depend on the energies of the theoretical single-particle states.
Hyperfine interaction, spin-orbit interaction and the shape of the
potential affect the positions of the single-particle states.
Although the fine structures are small in terms of the separation
between major shells, they are nonetheless important in determining
the locations of the dissociation temperatures.

With the additional threshold energies due to the angular momentum
selection rules, the dissociation temperatures $T_d$ in units of $T_c$
for $\psi'$, $\chi_{c2}$, $\chi_{c1}$, and $J/\psi$ are shifted to
0.50, 0.91, 0.90, and 0.99
, as listed in the first row of Table III.
The increase in the dissociation temperature is greatest for
$\chi_{c2}$ and least for $J/\psi$.

For the bottomium system, because the mass of the bottom quark is very
large, the mass difference between $B^*$ and $B$ and the centrifugal
barrier energies are small; the selection rules do not lead to
substantial changes of the thresholds. The shifts of the dissociation
temperatures are not as large as in the case of charmonium states.
The dissociation temperatures for various bottomium states are listed
in Table III. We confirm the general features of the results of Digal
$et~al.$ but there are also differences in the details as the
dissociation temperatures depend on the potential and interactions, as
well as on the selection rules.

With the temperature-dependent potential of Eqs.\
(\ref{eq:pot})-(\ref{eq:pot3}), all the dissociation temperatures of
heavy quarkonia are below $T_c$, except for $\chi_{b1}$, $\chi_{b2}$,
and $\Upsilon$.  The present analysis uses a potential valid in the
range $T\le T_c$ and cannot be used to determine dissociation
temperatures beyond $T_c$.

\section{Dissociation of quarkonium by thermalization}

Besides spontaneous dissociation, dissociation by interaction with
particles in the medium can also lead to dissociation.  In order to
bring out the salient features, it is useful to present simplifying
descriptions of these dissociation processes in terms of dissociation
by thermalization and dissociation by collisions.  A full description
will involve the interplay between the dynamics of thermalization and
the dynamics of different types of dissociation.

In the dissociation by thermalization, we consider a
two-step process.  In the first step, a thermal equilibrium can arise
from inelastic reactions of the type 
\begin{eqnarray}
\label{eq:equ}
h + (Q\bar Q)_{JLS} \to h' + (Q\bar Q)_{J'L'S'}
\end{eqnarray}
in which the collision of a hadron $h$ with a
quarkonium in state $JLS$ results in the excitation or de-excitation
of the quarkonium state.  If the heavy quarkonium reaches thermal
equilibrium with the medium, the occupation probabilities of the heavy
quarkonium state $\epsilon_i$ will be distributed according to the
Bose-Einstein distribution,
\begin{eqnarray}
\label{eq:be}
n_i={1\over \exp\{(\epsilon_i-\mu)/T\}-1}.
\end{eqnarray} 
If there is no dissociation of the quarkonium, then the sum of the
occupation numbers is
\begin{eqnarray}
\label{eq:normal}
n=\sum_i (2J_i+1)n_i =1,
\end{eqnarray}
where $2J_i+1$ is the spin degeneracy. This equation can be used to
determine the chemical potential $\mu$.  At temperature $T$, the
fraction of heavy quarkonium whose single-particle state energies lie
above their dissociation thresholds is
\begin{eqnarray}
f=\sum_i n_i\biggr | _{\epsilon_i \ge \epsilon_{i{\rm th}}}.
\end{eqnarray}

In the second step, after the quarkonium reaches thermal equilibrium
with the medium, there is a finite probability for the quarkonium
system to be found in excited states.  If these excited states lie
above their thresholds for spontaneous dissociation, the heavy
quarkonium system will have a finite probability to dissociate into
open charm or open bottom mesons.  

Previously, dissociation by thermalization was studied by Kharzeev,
McLerran, and Satz \cite{Kha95} with a temperature-independent
dissociation threshold and free-gas continuum states in a Boltzmann
distribution. They found that the dissociation by thermalization does
not provide a significant amount of $J/\psi$ dissociation at
temperature $T$=0.2 GeV.  Results in Sections \ref{sec:sps} and
\ref{sec:sps1} indicate however that the dissociation thresholds and
the positions of the single-particle states change with temperature.
Furthermore, the set of states $\epsilon_i$ included in Eqs.\
(\ref{eq:be}) and (\ref{eq:normal}) in the quarkonium thermal
equilibration process (\ref{eq:equ}) should consist only of bound and
resonance states \cite{Shl97}, because the thermal equilibration of
the heavy quarkonium $Q\bar Q$ takes place when $Q$ and $\bar Q$
remain in the vicinity of each other.  Free-gas continuum states off
the resonance represent $Q\bar q$ and $q\bar Q$ states whose spatial
amplitudes lie predominantly outside the quarkonium system and they do
not participate significantly in the quarkonium thermal equilibration
process (\ref{eq:equ}).

To study dissociation of heavy quarkonium by thermalization, we
evaluate the fraction $f$ of heavy quarkonium whose single-particle
energies lie above their dissociation thresholds, as a function of
temperature, using the single-particle states obtained in Sections
\ref{sec:sps} and \ref{sec:sps1}.  The results are shown in Fig.\ 6.
The fraction $f$ increases with temperature, due predominantly to the
change of the positions of the single-particle energies.  In Fig.\ 6,
a state label shown along the curves denotes the onset of spontaneous
dissociation of that state.  As the temperature approaches $T_c$, the
fraction lying above dissociation thresholds is quite large for
charmonium, and is considerably smaller for bottomium.

By the dissociation of the excited heavy quarkonium, the occupation
number is no longer constant and the rate of dissociation of the
quarkonium is
\begin{eqnarray}
{dn \over dt} =-\sum_i \lambda_i (2J_i+1) n_i\biggr 
| _{\epsilon_i \ge \epsilon_{i{\rm th}}},
\end{eqnarray} 
where $\lambda_i$ is the decay rate of the dissociating excited
quarkonium state $i$.  These dissociation rates for various excited
heavy quarkonium states have not been evaluated.  It will be
interesting to calculate these rates using explicit quark model wave
functions and the formalism discussed by Ackleh, Barnes and Swanson
\cite{Ack97}.

In order to appreciate the effect of dissociation by thermalization on
quarkonium survival probability, one can consider as an example the
case of placing a $J/\psi$ in a medium below the dissociation
temperature of $J/\psi$.  If the $J/\psi$ is not in thermal
equilibrium with the medium, then this $J/\psi$ system will be stable
against spontaneous dissociation. However, with the approach of
thermal equilibrium (by collision with particles in the medium) at a
temperature $T$, the initial $J/\psi$ system will evolve into a mixed
state with a probability distribution to populate different states of
the quarkonium system, some of which lie above their dissociation
thresholds at that temperature.  As a result, a fraction of the
charmonium system will dissociate into open charm mesons, even though
the temperature is below the $J/\psi$ dissociation temperature.  Thus,
a non-equilibrated $J/\psi$ that is stable in the medium may become
partially unstable against dissociation when it reaches thermal
equilibrium.  The dissociation probability of a quarkonium system
depends on its state of thermal equilibrium or non-equilibrium.

\vspace*{1.5cm}
\epsfxsize=300pt
\includegraphics{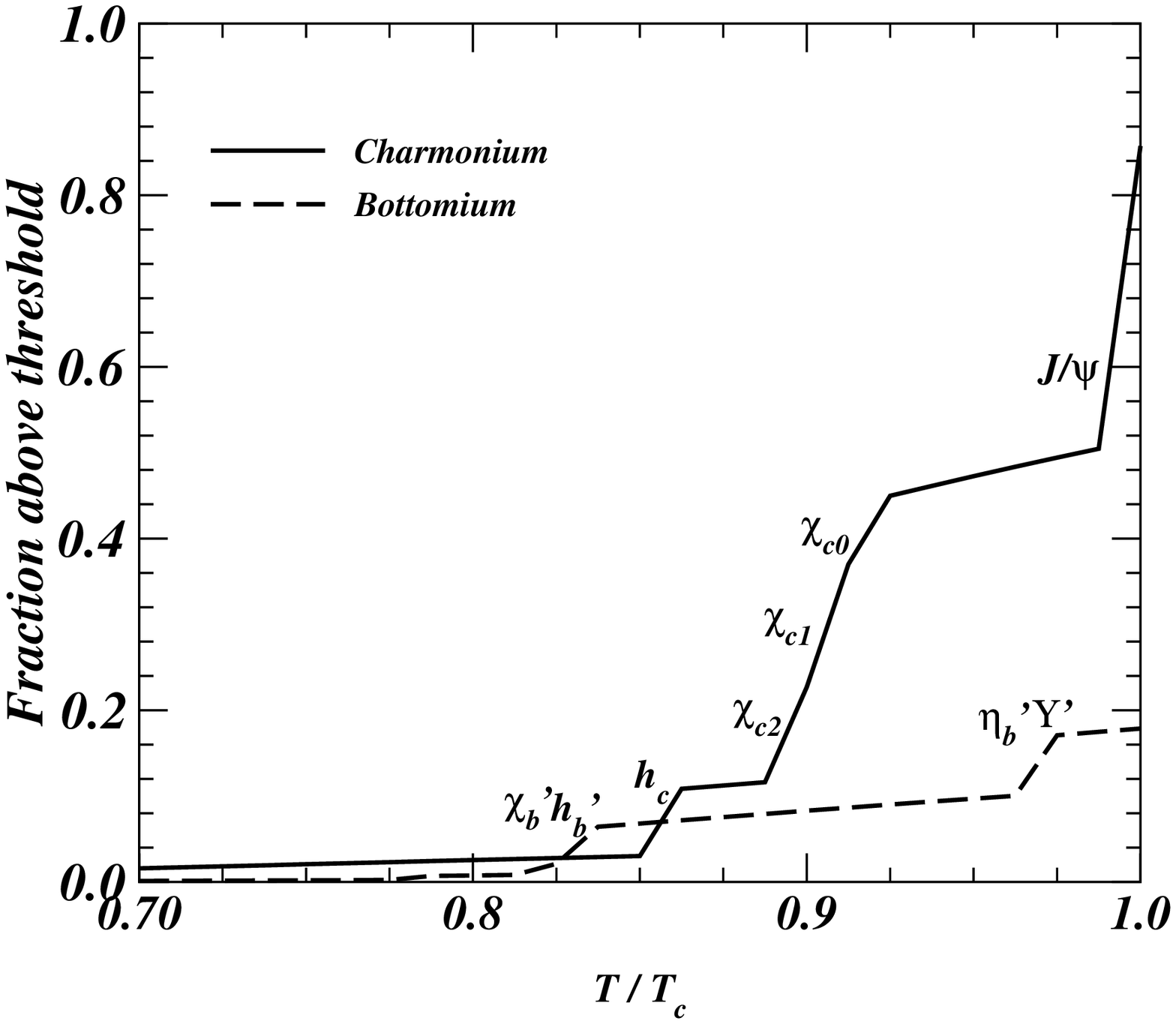}
\vspace*{+8.4cm}\hspace*{1cm}
\begin{minipage}[t]{12cm}
\noindent {\bf Fig.\ 6}.  {The fractions of charmonium and bottomium
lying above the dissociation threshold as a function of $T/T_c$.}
\end{minipage}
\vskip 4truemm
\noindent 

\section{Dissociation of heavy quarkonium by collision}

In the hadron thermal bath in which a heavy quarkonium is placed,
there will be hadrons which will collide with the heavy quarkonium.
Dissociation can occur as a result of these collisions.

Previously, we studied the dissociation of heavy quarkonium by
collision with light hadrons at $T=0$ \cite{Won00,Won00a,Bar00,Won01}.
We study in this section the dissociation of $J/\psi$ and $\Upsilon$
by collision with pions in a medium at temperature $T$.  The heavy
quarkonia under consideration can be part of a system in thermal
equilibrium with the medium.  They can also be non-equilibrated heavy
quarkonia introduced into the medium.  The kinetic energy of collision
would be of the order of the temperature of the hadron gas.  As the
temperature of the medium increases, the quarkonium single-particle
energy increases and the energy needed to dissociate the heavy
quarkonium decreases.  If the energy threshold for the dissociation of
the heavy quarkonium is comparable to the temperature of the hadron
gas, the dissociation cross section will be large.

We can calculate the dissociation cross sections of $J/\psi$ and
$\Upsilon$ in collision with $\pi$ as a function of the temperature
using the Barnes and Swanson model \cite{Bar92}.  The calculation
requires the energies and wave functions of the initial and final
meson states, as well as the interquark interaction which leads to the
dissociation.  For the interquark interaction, we generalize the
temperature-dependent Yukawa and exponential interaction in Eqs.\
(\ref{eq:pot}) and (\ref{eq:pot3}) to the form
\begin{eqnarray} 
V_{ij}={\bbox{\lambda}(i) \over 2}\cdot {\bbox{\lambda}(j) \over 2} \left
\{{\alpha_s e^{-\mu(T)r} \over r} - {3 b(T) \over 4\mu(T)} e^{-\mu (T)
r} 
- {8 \pi \alpha_s \over 3 m_i m_j } \bbox{s}_i \cdot \bbox{s}_j \left (
{d^3 \over \pi^{3/2} } \right ) e^{-d^2 r^2} \right \}.
\end{eqnarray}                                                                 
For an antiquark, the generator $\bbox{\lambda}/2$ is replaced by
$-\bbox{\lambda}^{T}/2$.  We have already calculated the
single-particle energies and wave functions of $J/\psi$ and $\Upsilon$
at various temperatures in Sections \ref{sec:sps} and \ref{sec:sps1}.  We
can calculate the wave functions of $\pi$, open charm, and open bottom
mesons using the above interaction.  The dissociation reactions with
the lowest final meson masses are $J/\psi + \pi \to D \bar D^*, D^*
\bar D, D^*\bar D^*$ and $\Upsilon + \pi \to B \bar B^*, B^* \bar B,
B^*\bar B^*$.  As the final states involves the $D^*$ and the $B^*$
while the above potential of (\ref{eq:pot})-(\ref{eq:pot3}) refers to
$D\bar D$ or $B\bar B$ at $r\to \infty$, we also need the mass
differences of $M(D^*)-M(D)$ and $M(B^*)-M(B)$, which we shall take
from the experimental masses at $T=0$.  According to the spectral
analysis by Hatsuda, the mass of pion does not change substantially as
a function of temperature (see Fig.\ 3 of Ref.\ \cite{Hat01}).  We
shall keep the mass of pion to be the same as in $T=0$ in this
calculation.

The reaction matrix element of $V_{ij}$ is a product of the color
matrix element, the flavor matrix element, the spin matrix elements,
and the spatial matrix elements, as discussed in detail in
\cite{Bar92} and \cite{Won01}.  With our Gaussian basis states, the
spatial matrix elements of $V_{ij}$ between the initial and the final
states can be reduced to the evaluation of Gaussian integrals of the
Fourier transform of the interaction potential, as shown in Eqs.\
(48)-(50) of Ref.\ \cite{Won01}.  For the above Yukawa and
exponential interactions, the corresponding integrals are
\begin{eqnarray}
\int d{\bbox{q}} ~ e^{- ( {\bbox{q}}-{\bbox{q}}_0)^2/2\beta^2 } V_{\rm
Yukawa}({\bbox{q}})
={4 \pi \alpha_s \over q_0}
 (\sqrt{2 \pi} \beta )^{3}
{\cal I}_{-1},
\end{eqnarray}                                                                 
\begin{eqnarray}
\int d{\bbox{q}} ~ e^{- ( {\bbox{q}}-{\bbox{q}}_0)^2/2\beta^2 } V_{\rm
exponential}({\bbox{q}})
={4 \pi \over q_0} {3b(T) \over
4 \mu(T)}  (\sqrt{2 \pi} \beta )^{3}
{\cal I}_{0},
\end{eqnarray}                                                                 
where
\begin{eqnarray} 
{\cal I}_n=\int_0^\infty r^{1+n}e^{-\beta^2 r^2/2-\mu r}\sin q_0 r \, dr.
\end{eqnarray}
The integrals ${\cal I}_{-1}$ and ${\cal I}_0$ are
\begin{eqnarray} 
{\cal I}_{-1}=-{i \over 2 \beta} \sqrt{{\pi \over 2}}
\left \{ \exp\{z_1^2\} {\rm erfc}(z_1) 
       - \exp\{z_2^2\} {\rm erfc}(z_2) \right \}
\end{eqnarray} 
and
\begin{eqnarray} 
{\cal I}_{0}={i \over\sqrt{ 2} \beta^2} \sqrt{{\pi \over 2}}
\left \{ z_1 \exp\{z_1^2\} {\rm erfc}(z_1) 
       - z_2 \exp\{z_2^2\} {\rm erfc}(z_2) \right \},
\end{eqnarray} 
where 
\begin{eqnarray}
z_1={\mu  - i q_0 \over \sqrt{2} \beta},
\end{eqnarray}                                                                 
and
\begin{eqnarray}
z_2={\mu + i q_0 \over \sqrt{2} \beta}.
\end{eqnarray}                                                                 

Using these results, the reaction matrix elements can be evaluated and
the dissociation cross section can be calculated, as in \cite{Won01}.
The sum of dissociation cross sections for $\pi+J/\psi \to D\bar D^*,
D^*\bar D, D^* \bar D^* $ are shown in Fig. 7 for different
temperatures $T/T_c$, as a function of the kinetic energy,
$E_{KE}=\sqrt{s}-M_{A}-M_{B}$ where $M_A$ and $M_B$ are the masses of
the colliding mesons.  In Fig.\ 8 we show similar total dissociation
cross sections for $\pi+\Upsilon \to B\bar B^*, B^*\bar B, B^* \bar
B^* $ for various temperatures, as a function of $E_{KE}$.  Each cross
section curve is the average of the results from the ``prior'' and
``post'' formalisms, as explained in Refs.\ \cite{Swa92,Won01}.

\vspace*{1.5cm}
\epsfxsize=300pt
\includegraphics{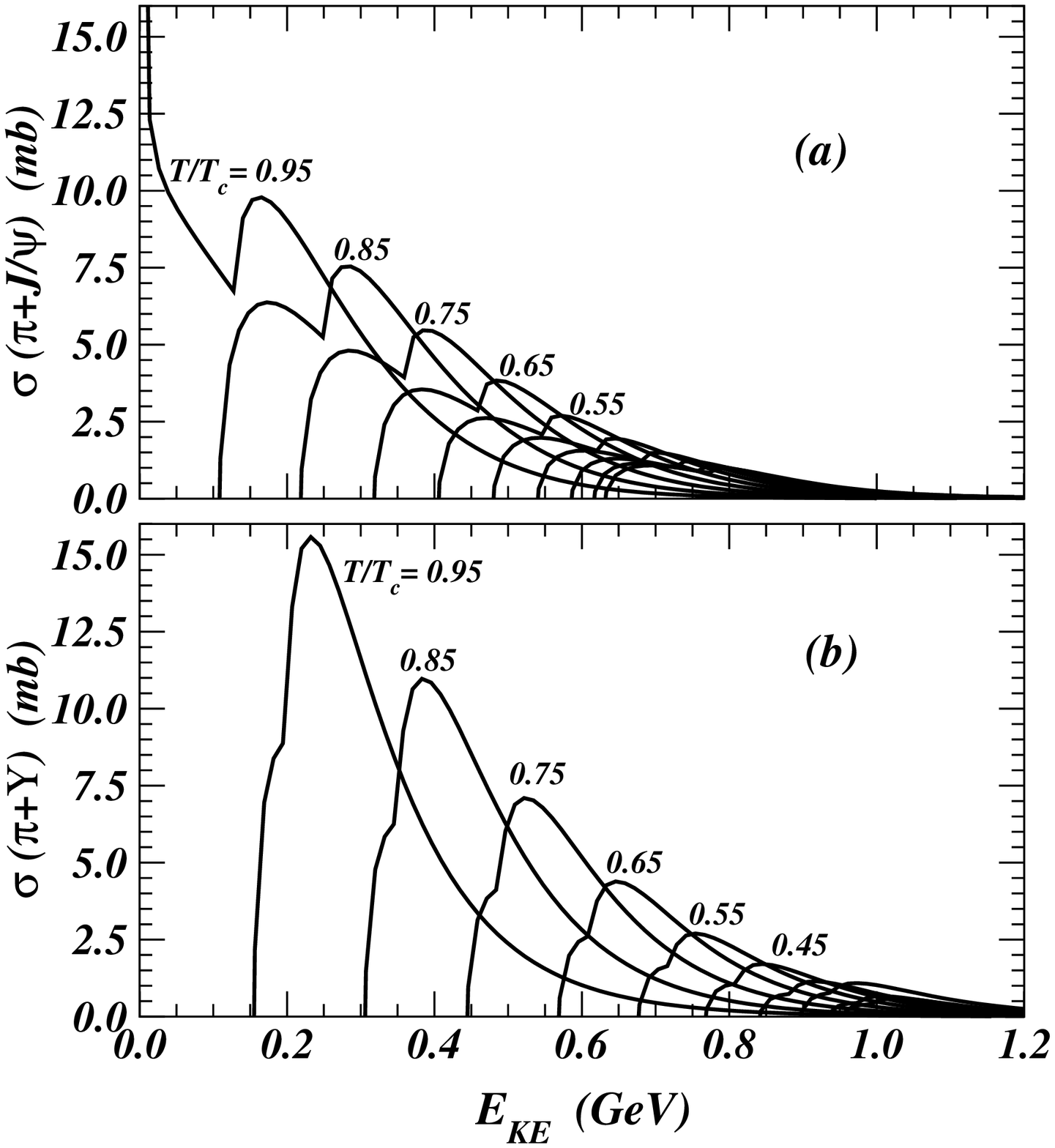}
\vspace*{+11.2cm}\hspace*{1cm}
\begin{minipage}[t]{12cm}
\noindent {\bf Fig.\ 7}.  {Total dissociation cross section of
$J/\psi$ in collision with $\pi$ for various temperatures
as a function of the kinetic energy
$E_{KE}$.} 
\end{minipage}
\vskip 4truemm
\noindent 

We observe in Fig.\ 7 that the dissociation cross sections increase
as the temperature increases. Such an increase arises from decreasing
threshold energies as the temperature increases.  At $T/T_c=0.95$ the
reaction $\pi + J/\psi \to D + \bar D^*$ is exothermic and the total
dissociation cross section diverges at $E_{KE}=0$.  There is another
cross section maximum of about 10 mb at $E_{KE}\sim 0.17$ GeV, which
arises from the $\pi+J/\psi \to D^* + \bar D^*$ reaction.  As
the temperature decreases, the cross section decreases.  At the
temperature of $T/T_c=0.7$, the maximum cross section decreases to about
5 mb at $E_{KE}\sim 0.3-0.5$ GeV.  At $T=0$, the maximum cross section
decreases to about 1 mb at $E_{KE}\sim 0.7$ GeV and the threshold is
at 0.64 GeV \cite{Won01}.

We present similar results for the dissociation of $\Upsilon$ in
collision with $\pi$ in Fig.\ 7$b$.  The reactions remain endothermic
even for $T$ close to $T_c$.  For $T/T_c=0.95$ the maximum cross section
of about 15 mb occurs at $E_{KE}\sim 0.2 $ GeV.  As the temperature
decreases, the maximum of the cross section decreases and moves to 
higher kinetic energy.  For $T/T_c=0.75$, the maximum cross section is
about 4 mb and is located at $E_{KE}=0.55$ GeV. At $T=0$, the maximum
cross section is about 0.6 mb at $E_{KE}\sim 1.04$ GeV and the
threshold is at about 1.00 GeV \cite{Won01}. The dissociation cross
section is a sensitive function of the threshold.

The results in Figs.\ 7  indicate that over a large range of
temperatures below the phase transition temperature, dissociation cross
sections of $J/\psi$ and $\Upsilon$ in collisions with $\pi$ are 
large.

\vspace*{1.5cm}
\epsfxsize=300pt
\includegraphics{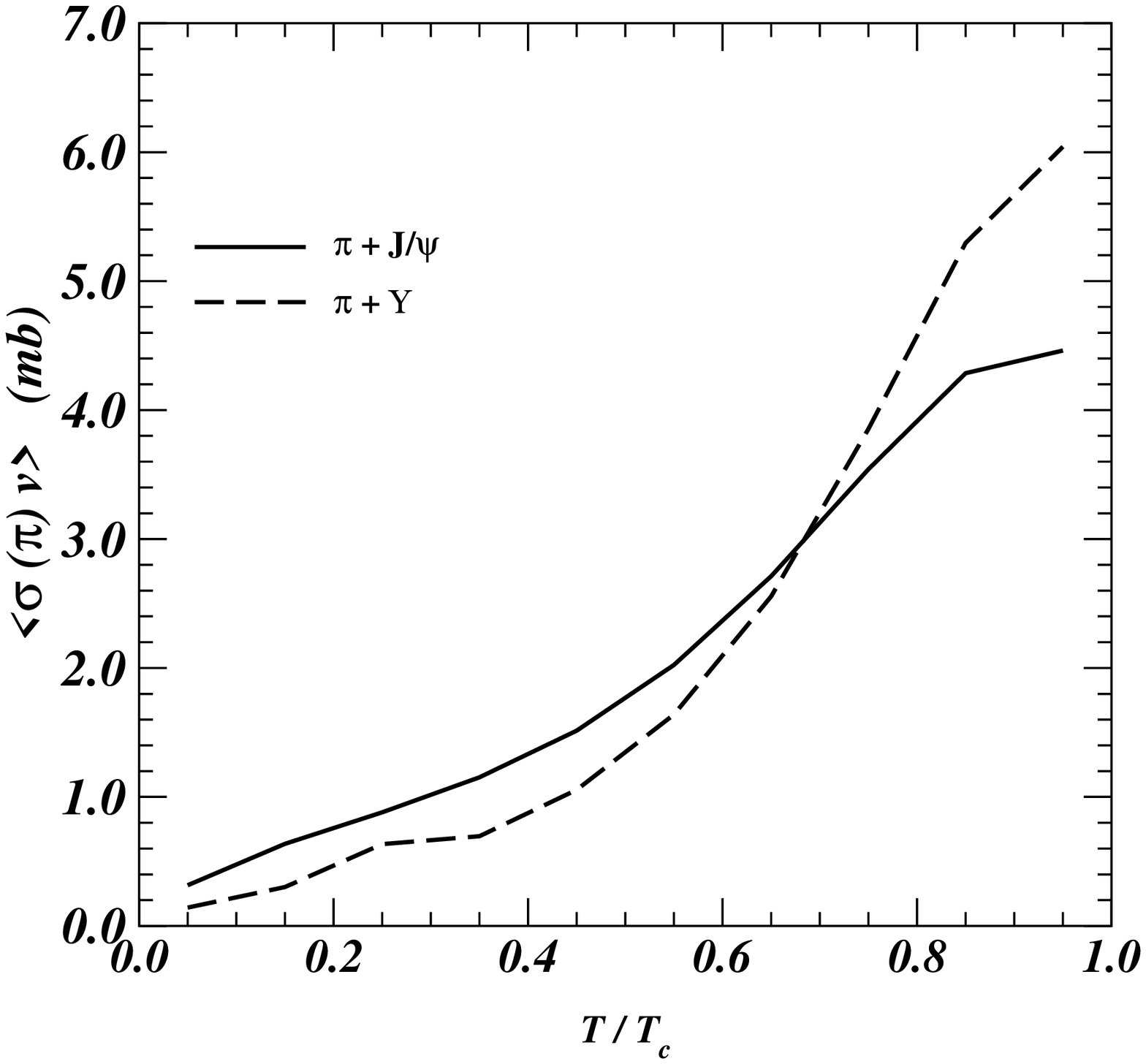}
\vspace*{+8.4cm}\hspace*{1cm}
\begin{minipage}[t]{12cm}
\noindent {\bf Fig.\ 8}.  {The average dissociation cross section of
$J/\psi$ and $\Upsilon$ in collision with $\pi$ as a function of
the temperature.
}
\end{minipage}
\vskip 4truemm
\noindent 

In a hadron gas, pions collide with the heavy quarkonium at different
energies.  We can get an idea of the energy-averaged magnitude of the
dissociation cross section by treating the pions as a Bose-Einstein
gas at temperature $T$.  In Figure 8, we show the quantity
$\langle \sigma v \rangle $ which is the product of the dissociation
cross section and the relative velocity, averaged over the energies of
the pions.  The quantity $\langle \sigma v \rangle$ is about 3 mb at
$T/T_c=0.70$ and rises to about 4.5 mb at $T/T_c=0.95$ where the value
of $T_c$ has been taken to be 0.175 GeV \cite{Kar00}.

\section {Dissociation of heavy quarkonium in collision with pions}

We can estimate the survival probability of a heavy quarkonium in a
hot pion gas in the presence of this type of collisional dissociation.
If we represent the survival probability $S$ by $\exp\{-I\}$, the
exponential factor $I$ is given by
\begin{eqnarray}
I= \int_{\tau_0}^{\tau_{\rm freeze}} \langle \sigma v \rangle (\tau)
\rho(\tau) d\tau
\end{eqnarray}
where $\sigma$ is the dissociation cross section, $v$ is the relative
velocity between $\pi$ and the heavy quarkonium, $\rho(\tau)$ is the
density of $\pi$ at the proper time $\tau$, and $\tau_0$ and
$\tau_{\rm freeze}$ are the initial proper time and the freeze-out
proper time respectively.  The quantity $\langle \sigma v \rangle$ in
Fig.\ 9 can be represented approximately by
\begin{eqnarray}
\langle \sigma v \rangle
\approx \langle \sigma v \rangle_c (T /T_c),
\end{eqnarray}
where $\langle \sigma v \rangle_c \sim 4.5 $ mb for $J/\psi$ or
$\Upsilon$.  Assuming a Bjorken type of expansion in which the density
$\rho(\tau)$ is proportional to $\tau^{-1}$ and $T(\tau)$ is
proportional to $\tau^{-1/3}$, we obtain
\begin{eqnarray}
I=\langle \sigma v \rangle_c \rho_0 \tau_0  
\left ({T_0\over T_c} \right ) 3 \left \{ 1 -
\left ({ \rho_{\rm freeze} \over \rho_0 }\right )^{1/3} \right \},
\end{eqnarray}
where $\rho_0$ is the pion density after all the pions are initially
formed and the heavy quarkonium is thermalized in the medium, $T_0$ is
the corresponding temperature, and $\rho_{\rm freeze}$ is the
freeze-out pion density.  Continuing the assumption of Bjorken
expansion, the initial pion density is given by
\begin{eqnarray}
\rho_0={dN \over \tau_0 dy {\cal A}},
\end{eqnarray}
where $dN/dy$ is the rapidity density of pions and $\cal A$ is the
overlapping area in the heavy-ion collision, which depend on
centrality.  We therefore obtain
\begin{eqnarray}
I= \langle \sigma v \rangle_c  {dN \over dy {\cal A}}  
\left ({T_0\over T_c} \right ) 
3 \left \{ 1 - \left ( {\rho_{\rm freeze}   \tau_0 {\cal A} 
\over dN/ dy } \right ) ^{1/3}\right \}.
\end{eqnarray}

To make some estimates on the survival probability of $J/\psi$ and
$\Upsilon$ in a hot hadron gas, we can take the quantity $\tau_0$ to
be (1 fm/c + 2 fm/c + $2R/\gamma$), where 1 fm/c is the formation time
for pions and heavy quarkonium, the additional 2 fm/c is the time for
the thermalization of hadrons, and $2R/\gamma$ is the approximate time
spread of the first and the last nucleon-nucleon collisions in a
nucleus-nucleus collision ($R$ is the radius of the colliding nucleus
and $\gamma$ is the relativistic factor in the C.M. frame).  We can
take $\rho_{\rm freeze}$, the freeze-out pion density, to be
$0.5$/fm$^3$, corresponding to an average pion separation of about 1.3
fm, and $T_0/T_c\sim 1$.

For the most central Pb-Pb collision at 158A GeV, $dN_{\rm ch}/dy\sim
450$ \cite{WA98}, and $dN/dy\sim 675$.  Then for this central
collision, $I=\langle \sigma v \rangle_c (2.40/{\rm fm}^2)$ and the
heavy quarkonium survival probability is $S=e^{-1.08}=0.34$.

For the most central Au-Au collision RHIC at $\sqrt{s_{NN}}=200$ GeV,
$dN_{\rm ch}/dy\sim 650$ \cite{Pho}, and $dN/dy\sim 975$.  Then for
this central collision, $I= \langle \sigma v \rangle_c (7.13/{\rm
fm}^2)$ and the heavy quarkonium survival probability is
$S=e^{-3.21}=0.04$.  There is a substantial absorption of both
$J/\psi$ and $\Upsilon$ by the hot pion gas at this temperature.

\section{Discussions and Conclusions}

We study the dissociation of a heavy quarkonium in high temperatures.
The temperature of the medium alters the gluon and quark fields and
changes the interaction between a heavy quark $Q$ and antiquark $\bar
Q$ placed in the medium.  We first examine the effects of the
temperature of the medium on the quarkonium single-particle states by
using a potential inferred from lattice gauge calculations and
evaluate the dissociation temperature at which the quarkonium
dissociate spontaneously.  We include spin-dependent interactions and
take into account the selection rules.  We confirm the general
features of the results of Digal $et~al.$ \cite{Dig01a} but there are
some differences in the dissociation temperatures, which depend on the
potential and interactions, as well as on the selection rules.  We
find that the selection rules change the dissociation temperatures
substantially for charmonia but only slightly for bottomia.

The results of Digal $et~al.$ and our work indicate that most, if not
all, of the dissociation temperatures of the charmonium states are
below $T_c$, but the dissociation temperatures of a number of low-lying
bottomium states lie above $T_c$ because of the strong binding of
these states.

The quarkonium placed in a medium can collide with particles in the
medium to reach thermal equilibrium with the medium.  A
heavy-quarkonium in thermal equilibrium can dissociate by
thermalization as there is a finite probability for the system to be
in an excited state lying above its dissociation threshold.  We find
that the fraction of the quarkonium lying above the threshold
increases with increasing temperatures and is quite large for
charmonium at $T$ slightly below $T_c$.

Dissociation of a heavy quarkonium can occur in collision with hadrons
at temperatures below the dissociation temperatures and the phase
transition temperature.  As the temperature increases, the quarkonium
single-particle energies increase, and the threshold energies for
collisional dissociation decrease.  As a consequence, the dissociation
cross sections increase as the temperature increases, reaching an
average of $\langle \sigma v \rangle \sim 3.0$ mb for the dissociation
of $J/\psi$ and $\Upsilon$ in collision with pions just below the
phase transition temperature.  We have estimated the survival
probability of $J/\psi$ and $\Upsilon$ in collision with pions in
central Pb-Pb collisions at SPS and RHIC energies and found the
survival probability to be small.  Dissociation of $J/\psi$ by
collisions with pions at high temperatures may be an important source
of the anomalous $J/\psi$ suppression in Pb-Pb collisions observed at
CERN \cite{Gon96}.

It has been suggested that collision of $J/\psi$ with ``comovers'' was
the source of the anomalous $J/\psi$ suppression in Pb-Pb collisions
\cite{Gav96,Cap97,Cas97}.  There were some uncertainties as the
dissociation cross section of the ``comover'' with $J/\psi$ was not
known and the dissociation cross sections for different hadrons are
very different.  The large threshold energy of 0.64 GeV for $J/\psi$
dissociation in collision with pions raised questions whether the
$\pi$-$J/\psi$ dissociation cross section could be large enough to
lead to the anomalous $J/\psi$ suppression.  The new information from
lattice gauge calculations of Karsch $et~al.$ \cite{Kar00} provides
valuable input to infer the in-medium dissociation energy of $J/\psi$.
The large cross sections for the in-medium dissociation in collision
with pions obtained here suggest that further microscopic
investigations of the $J/\psi$ dissociation in collision with hadrons
at high hadron temperatures will be of great interest.

The author would like to thank Prof. T. D. Lee for an illuminating
discussion on distinct signatures for the quark-gluon plasma.  The
author would like to thank Profs. Ted Barnes, C. M. Ko, K. F. Liu,
K. Redlich, R. Rapp, and S. Shlomo for stimulating discussions.  The
author thanks his colleagues Prof. S. Sorensen, Drs.  T. Awes,
V. Cianciolo, K. Read, F. Plasil, Xiao-Ming Xu, and Glenn Young for
helpful discussions.  The author thanks Profs. K. Haglin, D. Kharzeev,
and P. Preteczky for helpful communications.  This work was initiated
while the author was a summer visitor in the Nuclear Theory Group at
Brookhaven National Laboratory.  The author would like to thank
Prof. McLerran for his kind hospitality and for helpful discussions.
This research was supported by the Division of Nuclear Physics,
Department of Energy, under Contract No. DE-AC05-00OR22725 managed by
UT-Battelle, LLC.

\end{document}